\documentclass[aps,pra,reprint,superscriptaddress]{revtex4-2}

\usepackage{graphicx}
\usepackage{hyperref}
\usepackage{graphics}
\usepackage{color}

\begin{document}

\title{Role of the molecular environment in quenching the irradiation-driven fragmentation of Fe(CO)$_5$: a reactive molecular dynamics study}


\author{Benjamin Andreides}
\affiliation{J. Heyrovsk\'{y} Institute of Physical Chemistry, Czech Academy of Sciences, Dolej\v{s}kova 3, 18223 Prague, Czech Republic}

\author{Alexey V. Verkhovtsev}
\email{verkhovtsev@mbnexplorer.com}
\affiliation{MBN Research Center, Altenh\"oferallee 3, 60438 Frankfurt am Main, Germany}

\author{Juraj Fedor}
\email{juraj.fedor@jh-inst.cas.cz}
\affiliation{J. Heyrovsk\'{y} Institute of Physical Chemistry, Czech Academy of Sciences, Dolej\v{s}kova 3, 18223 Prague, Czech Republic}

\author{Andrey V. Solov'yov}
\affiliation{MBN Research Center, Altenh\"oferallee 3, 60438 Frankfurt am Main, Germany}

\date{\today}

\begin{abstract}
Irradiation-driven fragmentation and chemical transformations of molecular systems play a key role in nanofabrication processes where organometallic compounds break up due to the irradiation with focused particle beams. In this study, reactive molecular dynamics simulations have been performed to analyze the role of the molecular environment on the irradiation-induced fragmentation of molecular systems. As a case study, we consider the dissociative ionization of iron pentacarbonyl, Fe(CO)$_5$, a widely used precursor molecule for focused electron beam-induced deposition. In connection to recent experiments, the irradiation-induced fragmentation dynamics of an isolated Fe(CO)$_5^+$ molecule is studied and compared with that of a Fe(CO)$_5^+$ molecule embedded into an argon cluster. The appearance energies of different fragments of an isolated Fe(CO)$_5^+$ agree with the recent experimental data. For Fe(CO)$_5^+$ embedded into an argon cluster, the simulations reproduce the experimentally observed suppression of Fe(CO)$_5^+$ fragmentation and provide an atomistic-level understanding of this effect. Understanding irradiation-driven fragmentation patterns for molecular systems in environments facilitates the advancement of atomistic models of irradiation-induced chemistry processes involving complex molecular systems.
\end{abstract}

\maketitle

\section{Introduction}

Irradiation-driven chemistry (IDC) processes induced by the interaction of different types of radiation (X-rays, electrons, ion beams) with molecular systems are exploited in many modern and emerging technologies.
For instance, IDC processes play an important role in ion-beam radiotherapy \cite{Linz2012_IonBeams, schardt2010heavy} that exploits the ability of charged heavy particles to inactivate living cells due to the induction of complex DNA damage \cite{Nano-IBCT_book, Surdutovich_2014_EPJD.68.353, Nakano_2022_PNAS}.
IDC transformations in molecular films have been studied in relation to astrochemistry \cite{Mifsud_2021_EPJD.75.182, Mifsud_2022_PCCP.24.10974}. Such transformations occur during the formation of cosmic ices in the interstellar medium due to the interplay of molecular adsorption on a surface and surface irradiation \cite{Tielens_2013_RMP.85.1021}.

Electron-irradiation induced chemistry of organometallic molecules is central to focused electron beam induced deposition (FEBID) -- a technology for the controllable fabrication of complex nanostructures with nanometer resolution \cite{Utke_book_2012, DeTeresa-book2020, Winkler_2019_JAP_review, Huth_2021_JAP_review}.
In the FEBID process, electron-induced molecular fragmentation occurs via the dissociative ionization (DI), dissociative electron attachment (DEA), or neutral dissociation (ND) mechanisms, leading to the production of cationic, anionic or neutral fragments, respectively \cite{Thorman2015}.
Electron-induced decomposition of adsorbed precursor molecules releases organic ligands resulting in the clusterization of the precursor's metallic component on a surface.
The fundamental physicochemical phenomena that govern the formation, growth and composition of deposits grown by FEBID still need to be fully understood and are the subject of ongoing research \cite{Utke2022_CoordChemRev, Prosvetov2021_BJN, Prosvetov2022_PCCP}. Achieving this goal requires a concerted approach linking fundamental knowledge of electron-driven chemistry in FEBID \cite{Swiderek_2018_BJNano.9.1317} with rational design and synthesis of novel precursor molecules \cite{McElwee-White_2021_ACSApplMater}.

In recent years, much effort has been put into entangling the elementary processes, which lead to electron-induced cleavage of metal--ligand bonds; see review papers \cite{Thorman2015, Barth2020_JMaterChemC, Utke2022_CoordChemRev} and references therein. However, the vast majority of data on the electron-irradiation-induced processes involving FEBID precursor molecules is experimental. Typically, in experiments, a well-defined precursor target is crossed with a monochromatized electron beam, and the yields of reaction products are measured as a function of the projectile electron energy. Such experiments have been performed for precursor molecules in the gas phase \cite{Lacko_EPJD2015, Allan_PCCP2018, Ribar_EPJD2015}, those embedded in a cluster environment \cite{Lengyel2016_JPCC_1, Lengyel2016_JPCC_2, Lengyel2017_Beilstein, Lengyel2021}, and condensed on a surface as thin molecular films \cite{Massey_Sanche2015, Bilgilisoy_Fairbrother2021}; see also the recent reviews \cite{Barth2020_JMaterChemC, Utke2022_CoordChemRev}.

A detailed atomistic-level understanding of IDC processes (i.e., bond cleavage and further reactivity) in molecular systems can be developed through computational modeling.
A rigorous quantum-mechanical description of these processes, e.g. within time-dependent density functional theory (TDDFT), is feasible for relatively small molecular systems containing, at most, a few hundred atoms and evolving on the sub-picosecond timescale \cite{TDDFT_C2H6O2_radiolysis, TDDFT_H2O_tetramer, TDDFT_dR_fragmentation}.
Classical molecular dynamics (MD) represents an alternative theoretical framework for modeling complex molecular systems \cite{MBNbook_Springer_2017}.
However, standard classical MD does not account for the coupling of the system to incident radiation and does not describe irradiation-induced quantum transformations in the molecular system.

These limitations have been overcome by means of reactive classical MD \cite{Sushko2016_rCHARMM} and Irradiation Driven Molecular Dynamics (IDMD) \cite{Sushko2016_IDMD}, a novel methodology enabling the atomistic simulation of IDC processes in complex molecular systems.
These methodologies have been implemented into MBN Explorer \cite{Solovyov_2012_JCC_MBNExplorer}, the advanced software package for multiscale simulations of complex biomolecular, nano- and mesoscopic systems \cite{MBNbook_Springer_2017}.

\begin{sloppypar}
The IDMD methodology enables atomistic modeling of irradiation-driven processes, wherein the dynamics of molecular systems is accompanied by random, local quantum transformations (e.g. bond breakage via DI or DEA) induced in the system during its irradiation; see the examples in Refs.~\citenum{MBNbook_Springer_2017, Sushko2016_IDMD, DeVera_2020_SciRep, Prosvetov2021_BJN, Prosvetov2022_PCCP, DySoN_book_Springer_2022}.
Major dissociative transformations of irradiated molecular systems (such as molecular topology changes, redistribution of atomic partial charges, or alteration of interatomic interactions) are simulated by means of MD with reactive force fields, particularly the rCHARMM force field \cite{Sushko2016_rCHARMM} implemented in MBN Explorer.
\end{sloppypar}

In the present study, reactive MD simulations are carried out to analyze the effects of the molecular environment on the irradiation-induced fragmentation of molecular systems. As a case study, we consider electron-impact induced DI
of iron pentacarbonyl, Fe(CO)$_5$, one of the most common FEBID precursors for the fabrication of iron-based nanostructures \cite{Lukasczyk2008, Gavagnin2013, Gavagnin2014, deTeresa2016}.
The metal--ligand separation and the CO ligand dissociation processes are simulated using the reactive rCHARMM force field \cite{Sushko2016_rCHARMM} and quantified by analyzing appearance energies for different molecular fragments.
The role of the molecular environment is analyzed by comparing the irradiation-induced fragmentation dynamics of an isolated Fe(CO)$_5$ molecule with that of a Fe(CO)$_5$ molecule embedded into an argon cluster.

Two recent experimental studies provide a direct motivation for this work. Lacko et al. \cite{Lacko_EPJD2015} studied electron-impact induced DI of Fe(CO)$_5$ molecules in the gas phase. Different fragment ions corresponding to a sequential loss of individual CO ligands (down to a bare Fe$^+$ fragment) were observed, and the appearance energies of these ions were determined with high resolution. Lengyel et al. \cite{Lengyel2016_JPCC_2} studied the DI of Fe(CO)$_5$ picked up on argon clusters with a mean size of several hundred argon atoms. Strong suppression of ligand dissociation and a change in the Fe(CO)$_5$ ionization mechanism were observed \cite{Lengyel2016_JPCC_2}.
The simulation results reported in the present study agree with the results of these gas-phase and cluster-beam experiments.
For an isolated Fe(CO)$_5$ molecule, the main benchmark of the simulations -- appearance energies of the individual fragment ions -- is in good quantitative agreement with the experimental data. For Fe(CO)$_5$ embedded into an argon cluster, the simulations reproduce the experimentally observed suppression of Fe(CO)$_5$ fragmentation and provide an atomistic-level understanding of this effect.

The results reported in this study indicate the importance of understanding irradiation-driven fragmentation patterns for molecular systems in molecular environments. Such understanding may facilitate the advancement of atomistic models of irradiation-induced chemistry processes involving complex molecular systems.

\section{Computational methodology}
\label{sec:Methodology}

MD simulations of irradiation-driven fragmentation of Fe(CO)$_5$ have been performed by means of the MBN Explorer software package \cite{Solovyov_2012_JCC_MBNExplorer}. The MBN Studio toolkit \cite{Sushko_2019_MBNStudio} has been utilized to create the systems, prepare necessary input files and analyze simulation outputs.

\subsection{Interaction potentials}

Interatomic interactions for the Fe(CO)$_5$ molecule have been described using the reactive CHARMM (rCHARMM) force field introduced in Ref.~\citenum{Sushko2016_rCHARMM}. rCHARMM permits simulations of various molecular systems with the dynamically changing molecular topology \cite{Verkhovtsev_2017_EPJD.71.212, DeVera2019, Friis2020_JCC, Friis2021_PRE}, which is essential for modeling irradiation-driven transformations and chemistry. Examples of the application of rCHARMM \cite{Sushko2016_rCHARMM} to different molecular systems are summarized in a recent review \cite{Verkhovtsev_EPJD2021_IDMD} and a book \cite{DySoN_book_Springer_2022}.

The radial part of bonded interactions is described in rCHARMM by means of the Morse potential:
\begin{equation}
U^{{\rm bond}}(r_{ij}) = D_{ij} \left[ e^{-2\beta_{ij}(r_{ij} - r_0)} - 2e^{-\beta_{ij}(r_{ij} - r_0)} \right] .
\label{Eq. Morse}
\end{equation}
Here $D_{ij}$ is the dissociation energy of the bond between atoms $i$ and $j$, $r_0$ is the equilibrium bond length, and the parameter $\beta_{ij} = \sqrt{k_{ij}^{r} / D_{ij}}$ (with $k_{ij}^{r}$ being the bond force constant) determines the steepness of the potential. The bonded interactions are truncated at a user-defined cutoff distance beyond which the covalent bond gets broken and the molecular topology of the system changes.

The rupture of covalent bonds in the course of simulation employs the following reactive potential for valence angles \cite{Sushko2016_rCHARMM}:
\begin{equation}
U^{{\rm angle}}(\theta_{ijk}) =
2 k^\theta_{ijk} \, \sigma(r_{ij}) \, \sigma(r_{jk}) \left[ 1 - \cos(\theta_{ijk}-\theta_0 )  \right] ,
\label{Eq. Angles}
\end{equation}
where $\theta_0$ is the equilibrium angle formed by a triplet of atoms $i$, $j$ and $k$; $k^{\theta}$ is the angle force constant; and the function
\begin{equation}
\sigma(r_{ij}) = \frac{1}{2} \left[1-\tanh(\beta_{ij}(r_{ij}-r_{ij}^*))  \right]
\label{Eq. Rupture_param}
\end{equation}
describes the effect of bond breakage, see Ref.~\citenum{Sushko2016_rCHARMM} for the details. The parameter $r_{ij}^*$ in Eq.~(\ref{Eq. Rupture_param}) is given by
\begin{equation}
r_{ij}^* = \frac12 \left( R^{{\rm vdW}}_{ij}+r_0 \right) ,
\end{equation}
where $r_0$ is the equilibrium distance between two atoms involved in the angular interaction and $R^{{\rm vdW}}_{ij}$ is the sum of the van der Waals radii for those atoms.

\begin{table*}[t!]
\centering
\caption{Covalent bonded and angular interaction parameters for a Fe(CO)$_5^+$ molecule employed in this study.}
\begin{tabular}{p{3.5cm}p{2cm}p{1.5cm}p{1.5cm}p{3.5cm}}
\hline
bond type      &   $r_0$~(\AA)   &  \multicolumn{2}{c}{$D_{ij}$}
  &  $k_{ij}^{r}$~(kcal/mol \AA$^{-2}$)  \\
   &     & (kcal/mol)  & (eV) & \\
\hline
Fe -- C$_{\rm ax}$      &  1.88  &   37.1  &  1.61  &   111.3  \\
Fe -- C$_{\rm eq}$      &  1.90  &   25.6  &  1.11  &    78.2  \\
C$_{\rm ax / eq }$ -- O &  1.12  &  227.6  &  9.87  &  1564.3  \\
\hline
\hline
angle type  &  $\theta_0$~(deg.)  & \multicolumn{2}{c}{$k_{ijk}^{\rm {\theta}}$~(kcal/mol rad$^{-2}$)}   \\
\hline
C$_{\rm ax}$ -- Fe -- C$_{\rm ax}$   &  180   &  76.4  \\
C$_{\rm eq}$ -- Fe -- C$_{\rm eq}$   &  120   &  76.4  \\
C$_{\rm ax}$ -- Fe -- C$_{\rm eq}$   &   90   &  76.4  \\
Fe -- C$_{\rm ax / eq}$ -- O         &  180   &  28.0  \\
\hline
\end{tabular}
\label{Table:CovBonds}
\end{table*}

\begin{table}[t!]
\caption{Parameters of the Lennard-Jones potential, Eq.~(\ref{Eq. Lennard-Jones}), describing the van der Waals interaction between atoms of the Fe(CO)$_5^+$ molecule and argon atoms.}
\centering
\begin{tabular}{p{1.5cm}p{2.5cm}p{2.5cm}p{1cm}}
%
\hline
Atom	& $\varepsilon$ (eV) & $r^{{\rm min}}/2$~(\AA) & Ref. \\
		\hline
	Fe    &  0.0024   &  2.27   &  \citenum{Mayo1990}         \\
	C     &  0.0041   &  1.95   &  \citenum{Mayo1990}         \\
	O     &  0.0042   &  1.76   &  \citenum{Mayo1990}         \\
	Ar    &  0.0103   &  1.91   &  \citenum{Talu_2001_LJ_Ar}  \\
		\hline
\end{tabular}
\label{Table:van_der_Waals}
\end{table}


In the present study, the results of MD simulations are compared with the experimental results on electron-impact induced dissociative ionization \cite{Lacko_EPJD2015, Lengyel2016_JPCC_2}. As such, a singly charged parent molecule, Fe(CO)$_5^+$, is considered in the simulations.
Within the utilized computational methodology based on classical reactive MD simulations, it is assumed that the positively charged Fe(CO)$_5^+$ molecule is in its ground electronic state. The electron impact ionization process studied experimentally in Refs.~\citenum{Lacko_EPJD2015, Lengyel2016_JPCC_2} can lead to electron removal from different molecular orbitals, and the cation can thus be formed in many different initial electronic states (with holes in the corresponding orbitals). Such quantum processes occurring at the initial time instant are not considered within the classical MD framework. However, it is well established that the internal conversion of excited states, e.g. via conical intersections, populates the cation's electronic ground state \cite{li20, graves21}. The conversion to the ground state is a fast process, typically proceeding within tens of femtoseconds \cite{rankovic_novec19}. The excess energy, initially stored in the electronic degrees of freedom, is thus transferred to the vibrational degrees of freedom in the ground state. This represents the starting point of our simulations and justifies the use of classical reactive MD to characterize the fragmentation patterns.

\begin{figure}[ht!]
\includegraphics[width=0.3\textwidth]{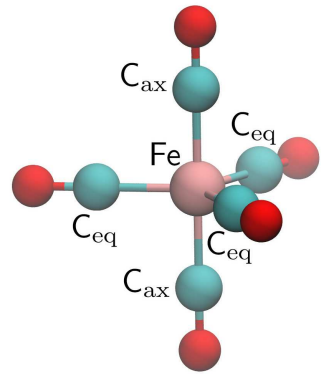}
\caption{Optimized geometry of a Fe(CO)$_5^+$ molecule considered in the present study. The optimization calculation has been
performed by means of MBN Explorer using the interatomic
potentials given by Eqs.~(\ref{Eq. Morse})--(\ref{Eq. Lennard-Jones}). Different atom types are indicated. The corresponding bonded and angular interactions are listed in Table~\ref{Table:CovBonds}.}
\label{fig:FeCO5_structure}
\end{figure}


A Fe(CO)$_{5}$ molecule has a $D_{3h}$-symmetric structure in the gas phase \cite{Portius_2019_FeCO5_structure} with two axial (``ax'') CO groups lying on the main symmetry axis of the molecule and three equatorial (``eq'') CO groups lying in the plane perpendicular to the main symmetry axis.
Parameters of the rCHARMM force field for Fe(CO)$_5^+$ have been determined from DFT calculations performed using Gaussian 09 software \cite{Gaussian09}. The optimized geometry of Fe(CO)$_5^+$ is shown in Fig.~\ref{fig:FeCO5_structure}.
Atomic partial charges for singly charged and neutral Fe(CO)$_5$ molecules were obtained through the natural bond orbital analysis \cite{Gaussian09}.
The parameters of the bonded and angular interactions for Fe(CO)$_5^+$ are listed in Table~\ref{Table:CovBonds}. These parameters were also employed in the recent study \cite{Prosvetov2022_PCCP} to simulate, by means of IDMD, the formation and growth of iron-containing nanostructures during the FEBID of Fe(CO)$_5$.


Non-bonded van der Waals interactions between atoms of the Fe(CO)$_5^+$ molecule have been described by means of the Lennard-Jones potential:
\begin{equation}
U_{{\rm LJ}}(r_{ij}) =
\varepsilon_{ij} \,
\left [
\left (\frac{r^{{\rm min}}}{r_{ij}} \right )^{12} -
2\left (\frac{r^{{\rm min}}}{r_{ij}}\right )^6
\right ] ,
\label{Eq. Lennard-Jones}
\end{equation}
where $\varepsilon_{ij}=\sqrt{\varepsilon_i \, \varepsilon_j}$ and $r^{{\rm min}} = (r^{{\rm min}}_i+r^{{\rm min}}_j)/2$.
The corresponding parameters are listed in Table~\ref{Table:van_der_Waals}.

\subsection{Fragmentation of an isolated Fe(CO)$_5$ molecule}
\label{sec:theor_gas}

Simulations of electron-impact induced fragmentation of an isolated iron pentacarbonyl molecule have followed the methodology from Ref.~\citenum{DeVera2019}. In the cited study, a model for irradiation-induced molecular fragmentation was developed on the basis of reactive MD simulations of W(CO)$_6$ fragmentation.
Two stages of the fragmentation process were considered within the model. \textit{(i)}~The localized energy deposition into a specific covalent bond immediately after the ionization or electron attachment processes. These processes happen on a sub-femtosecond scale and leave the molecular system in an excited electronic state. An excitation involving an antibonding molecular orbital evolves through the cleavage of a specific bond on the femtosecond timescale. \textit{(ii)}~Energy transfer into the system’s vibrational degrees of freedom via the electron-phonon coupling mechanism \cite{Gerchikov_2000_JPB.33.4905}. This process happens on a picosecond timescale after the collision, and the subsequent molecular fragmentation may last up to microseconds.

Within the framework of classical reactive MD, we have simulated both the cleavage of individual covalent bonds and energy redistribution over all the molecular degrees of freedom. Both processes result in an increase in the molecule's internal energy after the energy deposition. The internal energy increase is treated as an initial increase in the kinetic energy of atoms. For simulations of the first fragmentation mechanism, the amount of energy $E$ remaining in the system after ionization (i.e., excess energy over the first ionization potential) has been deposited locally into a specific covalent bond of the target and converted into the kinetic energy of the two atoms forming the bond. Velocities of these atoms have been defined to obey the total energy and momentum conservation laws:
\begin{equation}
{\bf v}_1 =  \frac{ \sqrt{2\mu E } }{m_1}{\bf u} \ , \qquad
{\bf v}_2 = -\frac{ \sqrt{2\mu E } }{m_2}{\bf u} \ .
\label{eq:modtwo}
\end{equation}
Here $m_1$, $m_2$ and $\mu = m_1 m_2/(m_1 + m_2)$ are, respectively, masses and the reduced mass of the atoms forming the bond, and ${\bf u}$ is a unit vector defining the direction of the relative velocity of these atoms upon bond cleavage.

The thermal mechanism of fragmentation corresponds to a statistical distribution of the deposited energy over all the degrees of freedom of the target. In this case, equilibrium velocities of atoms corresponding to a given temperature, $v_{i}^{\rm eq}$,
have been scaled by a factor $\alpha$ depending on the amount of deposited energy. The kinetic energy of $N$ atoms is then given by:
\begin{equation}
\sum_i^{N} \frac{1}{2}m_i (\alpha \, v_i^{\rm eq})^2
=
\frac32 N k_{\rm B}T + E .
\label{eq:modall}
\end{equation}
The first term on the right-hand side of Eq.~(\ref{eq:modall}) is the kinetic energy of the atoms at the equilibrium temperature $T$, with $k_{\rm B}$ being the Boltzmann's constant. The second term on the right-hand side is the excess energy deposited in the molecule during the collision.

The simulations of electron-impact induced dissociative ionization of an isolated Fe(CO)$_5$ molecule have been performed based on the following computational protocol.
First, the geometry of the molecule was optimized by means of MBN Explorer using the parameters listed in Tables~\ref{Table:CovBonds} and \ref{Table:van_der_Waals}. Then, the molecule was thermalized at $T = 300$~K; ten independent MD simulations of 1~ns duration each were performed. The simulations were performed using the Langevin thermostat with a damping time of 0.2~ps. In each trajectory, atomic coordinates and velocities were recorded every 100~ps. The simulated trajectories were used to generate a series of initial geometries and velocity distributions for the simulation of the fragmentation process.

The reactive MD simulations of Fe(CO)$_5^+$ fragmentation have been performed over 100~ns in a large simulation box with a side length of 200~\AA. The simulations used the integration time step of 0.1~fs, and no thermostat was employed.
Four thousand constant-energy simulations were conducted for different values of the excess energy $E$ ranging from 0 to $\sim$15.2~eV (350~kcal/mol). The largest value of $E$ considered here is about ten times larger than the dissociation energy for a Fe--C bond (see Table~\ref{Table:CovBonds}), which enables the simulation of multiple Fe--C bond breaks. The amount of energy $E$ has been varied from 0 to $\sim$10.8~eV (250~kcal/mol) in steps of $\sim$0.55~eV (12.5~kcal/mol). At higher $E$ values, a larger increment of $\sim$1.1~eV (25~kcal/mol) was considered. Molecular fragments produced at the end of 100-ns-long simulations were analyzed. The corresponding fragment appearance energies were evaluated from this analysis and compared with experimental data \cite{Lacko_EPJD2015}.

\subsection{Fragmentation of Fe(CO)$_5$ embedded into argon cluster}
\label{sec:theor_cluster}

The simulations of electron-impact induced fragmentation of a Fe(CO)$_5^+$ molecule embedded into an argon cluster have been set up according to the experimental parameters from Ref.~\citenum{Lengyel2016_JPCC_2}. In the cited study, the mixed Fe(CO)$_5$@Ar compounds were prepared by passing the argon cluster beam via a pick-up cell filled with the vapor of iron pentacarbonyl; the resulting heterogeneous clusters were ionized by the electron impact.

The process of Fe(CO)$_5^+$ pick-up by argon clusters has been simulated by means of classical MD. First, a spherical argon cluster with a radius of 1.3~nm, containing 230 atoms, has been created using the modeller plug-in of MBN Studio \cite{Sushko_2019_MBNStudio}.
The cluster has been thermalized at 40~K over 1~ns.
The interaction between argon atoms has been described using the Lennard-Jones potential, Eq.~(\ref{Eq. Lennard-Jones}), with the parameters listed in Table~\ref{Table:van_der_Waals}.
The simulations of Fe(CO)$_5^+$ pick-up on argon have been set up according to the experimental conditions of Ref.~\citenum{Lengyel2016_JPCC_2}.
A single Fe(CO)$_5^+$ molecule thermalized at 300~K collided with a cold argon cluster thermalized at 40~K. The collision velocity was set equal to 4.9~\AA/ps (490~m/s), corresponding to an average collision velocity in the experiment \cite{Lengyel2016_JPCC_2}. The simulations have been performed for 10~ns.



The resulting geometry of a heterogeneous Fe(CO)$_5^+$@Ar cluster has been used as an input for the simulation of fragmentation of Fe(CO)$_5^+$ embedded in the cluster. The simulation protocol is similar to that described above in
Section ``Fragmentation of an isolated Fe(CO)$_5$ molecule''.
%
An amount of energy $E$ ranging from 0 to $\sim$21.7~eV (500~kcal/mol) was deposited into different Fe--C and C--O bonds of the molecule. We have considered an increment of $\sim$2.2~eV (50~kcal/mol) over the whole energy range considered. In addition, the parameter $E$ was varied in steps of $\sim$0.4~eV (10~kcal/mol) in the range $E \approx 4.3 - 8.7$~eV (from 100 to 200 kcal/mol). The selected energy range corresponds to the range of appearance energies reported in the experiment \cite{Lengyel2016_JPCC_2}. The chosen increment of $\sim$0.4~eV corresponds to the experimental resolution reported in the cited study.
2800 MD simulations employing the rCHARMM force field have been carried out. The duration of each simulation was set to 25~ns with a time step of 0.1~fs.


\section{Results and discussion}
\label{sec:Results}

\subsection{Fragmentation of isolated iron pentacarbonyl}
\label{sec:Results_isolated_FeCO5}

\begin{figure}[t!]
\includegraphics[width=0.42\textwidth]{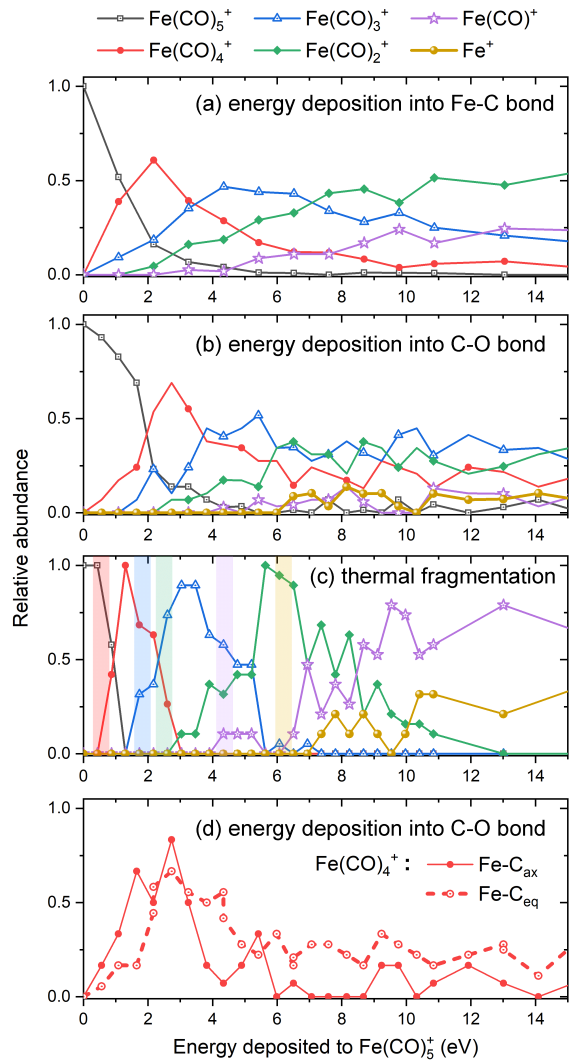}
\caption{Panels \textbf{(a)-(c)}: Relative abundance of Fe(CO)$_{5-n}^+$ ($n=0 - 5$) fragments produced after the electron impact induced fragmentation of an isolated Fe(CO)$_5^+$ molecule as a function of energy $E$ deposited to the molecule as a result of the collision.
Panels~(a) and (b) describe the case of the localized energy deposition into a Fe--C bond and a C--O bond, respectively.
Panel~(c) describes the thermal mechanism of fragmentation where the energy is distributed over all degrees of freedom of the molecule.
Vertical color bars in panel~(c) indicate the experimental appearance energies of Fe(CO)$_{5-n}^+$ ($n=1 - 5$) fragments reported by Lacko \textit{et al.}~\cite{Lacko_EPJD2015}, the width of these bars corresponds to the experimental uncertainty.
Legend corresponds to the data plotted in panels (a), (b) and (c).
Panel~\textbf{(d)}: Relative abundances of Fe(CO)$_4^+$ fragments produced by the removal of one CO ligand. Solid line corresponds to the localized energy deposition in a Fe--C$_{\rm ax}$ coordinated C--O bond; dashed line describes the case of energy deposition in a Fe--C$_{\rm eq}$ coordinated C--O bond.}
\label{fig:gas_phase}
\end{figure}

The main outcome of the performed simulations is the fragmentation patterns, that is, the abundance of different Fe(CO)$_{5-n}^+$ ($n=0 - 5$) ionic products as a function of energy $E$ deposited to the parent Fe(CO)$_5^+$ ion. Figure~\ref{fig:gas_phase}(a-c) shows this abundance for
the case when the excess energy has been deposited locally into one of Fe--C bonds or one of C--O bonds (panels~(a) and (b), respectively) and when the energy has been redistributed over all the degrees of freedom of the molecule (panel~(c)).

The abundance distributions for fragments produced after the localized energy deposition and the energy redistributed over all the molecular degrees of freedom have common features. In particular, each fragment ion has specific appearance energy; at larger values of the excess energy $E$, the abundance distribution reaches a maximum and then decreases with a further increase of $E$.
The decrease is caused by the fact that if the ion is too ``hot'' a larger number of CO ligands are evaporated. Note that the horizontal axis in Fig.~\ref{fig:gas_phase}, the energy deposited to the Fe(CO)$_5^+$ parent molecule, can be converted into the energy of projectile electrons by adding the ionization energy of a Fe(CO)$_5$ molecule, $I = 8.45$~eV, to this value.
The vertical color bars in Fig.~\ref{fig:gas_phase}(c) illustrate the appearance energies of Fe(CO)$_{5-n}^+$ ($n=1 - 5$) fragments determined in mass-spectroscopic experiments by Lacko et al.~\cite{Lacko_EPJD2015}, which have been converted into the excess energy deposited to Fe(CO)$_5^+$. Figure~\ref{fig:gas_phase}(c) indicates a very good agreement of the fragment appearance energies evaluated from the present simulations and the corresponding experimental values.

The qualitative similarity of the fragmentation patterns as a result of the different scenarios of energy deposition into the target molecule points out to strong intramolecular vibrational redistribution (IVR). When the energy has been deposited to a specific bond (either Fe--C or C--O), it is transferred to the neighboring atoms and redistributed among the vibrational degrees of freedom. Thus, the subsequent dissociation dynamics proceeds similarly to the case when the energy has been distributed over all the degrees of freedom at the beginning of the simulation, see Fig.~\ref{fig:gas_phase}(c).

Still, there are several quantitative differences between the simulated fragmentation patterns. First, the lowest fragment appearance energies correspond to the localized energy deposition into a Fe--C bond (Fig.~\ref{fig:gas_phase}(a)), followed by the case when the energy is deposited into a C--O bond (Fig.~\ref{fig:gas_phase}(b)), and the highest appearance energies correspond to the thermal mechanism of fragmentation where the energy is distributed over  all degrees of freedom of the molecule (Fig.~\ref{fig:gas_phase}(c)). The second difference concerns the width of fragment abundance distributions. The narrowest distributions correspond to the case of thermal fragmentation; somewhat broader distributions result from the energy deposited into a C--O bond, and the broadest distributions arise when the energy is deposited to a Fe--C bond.

As detailed in the Computational Methodology, section ``Fragmentation of an isolated Fe(CO)$_5$ molecule'',
dissociation energies for the Fe$_{\rm ax}$--C and Fe$_{\rm eq}$--C bonds differ by 0.5~eV (see Table~\ref{Table:CovBonds}). We have explored whether the resulting fragmentation pattern depends on the localized energy deposition to the C--O bonds coordinated to the different sites. The only detectable difference concerns the formation of the Fe(CO)$_4^+$ fragment, i.e. the loss of one ligand, see Figure~\ref{fig:gas_phase}(d). Abundances of other fragments are very similar for the two considered cases.
The efficient IVR leads to energy distribution over the entire molecule and to the fact that the energy deposition into a C--O bond coordinated either to a C$_{\rm ax}$ site or a C$_{\rm eq}$ site (where it is bound much more weakly, see Table~\ref{Table:CovBonds}) plays a minor role in the dissociation process.


It should be noted that the appearance energies of fragments are, in principle, the only quantities that can be directly compared to the mass-spectroscopic experimental data. In the electron impact ionization process,
\begin{equation}
e^- + \mbox{Fe(CO)}_5 \to \mbox{Fe(CO)}_5^+ + 2e^-,
\end{equation}
the two outgoing electrons carry away a certain fraction of the incident electron energy. The excess energy stored in the Fe(CO)$_5^+$ molecule after the collision
(the parameter which we control in the simulations) depends on the kinetic energy of these electrons. An $(e,2e)$ type of experiment, where the energies of all involved electrons are monitored in coincidence with the ionic fragmentation pattern~\cite{coplan94}, would be perfect for the comparison with the outcomes of present simulations. Unfortunately, we are not aware of any published data for $(e,2e)$ experiments with iron pentacarbonyl.

\subsection{Fragmentation of iron pentacarbonyl embedded into argon cluster}
\label{sec:Results_FeCO5@Ar}

In the experiments described in Ref.~\citenum{Lengyel2016_JPCC_2}, iron pentacarbonyl molecules were picked up by an argon cluster beam. A pertinent question in such pick-up experiments is the structure of the resulting heterogeneous Fe(CO)$_5$@Ar system \cite{fedor_pickup11, lengyel12, pysanenko15, farnik21}, i.e. do Fe(CO)$_5$ molecules stay on the surface of the cluster, or do they penetrate it and are embedded inside?
This information is important for determining the further reactivity of the picked-up species and their interaction with incident electrons, which will depend on whether the molecules are covered by rare gas atoms or located on a cluster surface.
This information cannot be obtained directly in experiments.

To address this question, we have simulated the collision of neutral and singly charged Fe(CO)$_5$ molecules with an Ar$_{230}$ cluster.
The simulations of Fe(CO)$_5$ pick-up on argon have been set up according to the experimental conditions of Ref.~\citenum{Lengyel2016_JPCC_2}.
%
%
As described above, a Fe(CO)$_5$ molecule collided with an argon cluster with the velocity of 490~m/s, corresponding to an average collision velocity in the experiment \cite{Lengyel2016_JPCC_2}.
Three different collision geometries have been considered: (i) a central hit corresponding to the zero impact parameter, $b=0$, (ii) a ``lateral'' hit with the impact parameter smaller than the radius of the cluster, $b < R$, and (iii) an ``orbital'' hit with $b \sim R$. By considering different collision geometries, we have explored whether the molecule stays on top of the cluster or penetrates its interior region after the collision.
Collision-induced evaporation of some loosely bound argon atoms has been observed in the performed simulations. As a result, after the collision, the Fe(CO)$_5^+$@Ar compound contained about 200 argon atoms.

\begin{figure}[t]
\includegraphics[width=0.48\textwidth]{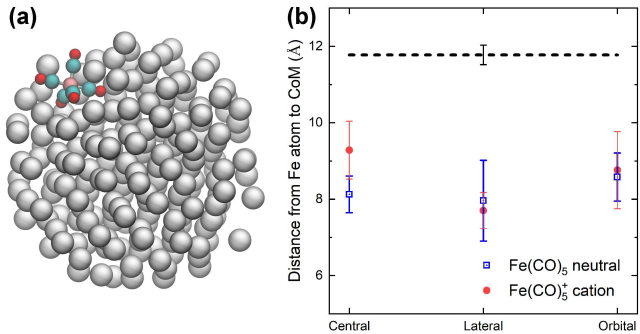}
\caption{Panel~\textbf{(a)}: An exemplary snapshot of the Fe(CO)$_5$@Ar cluster formed after the collision of a Fe(CO)$_5$ molecule with the argon cluster. Panel~\textbf{(b)}: Average distance between the iron atom of Fe(CO)$_5$ and the center of mass (CoM) of the cluster. The dashed line corresponds to the average radius of the Fe(CO)$_5$@Ar cluster at the end of 10-ns long simulations. Five independent MD simulations of Fe(CO)$_5$--Ar collisions have been performed for each charge state of Fe(CO)$_5$ and each collision geometry; error bars indicate the corresponding standard deviation.}
\label{fig:pickup}
\end{figure}

Figure~\ref{fig:pickup}(a) shows a typical structure of the Fe(CO)$_5$@Ar cluster at the end of a 10~ns long simulation of the molecule pick-up process. Five independent trajectories have been simulated for each geometry of the molecule--cluster collision; the results of this analysis are shown in Figure~\ref{fig:pickup}(b). The figure shows that the Fe(CO)$_5$ molecule is embedded into the cluster but stays relatively close to the cluster surface. The average distance between the iron atom and the cluster surface varies between 3 and 4~\AA, which is comparable with the distance between the Fe and O atoms in the Fe(CO)$_5$ molecule.

The specific pick-up process in the above-described experiment proceeds via the argon cluster collision with a neutral Fe(CO)$_5$ molecule.
In the present study, complementary simulations have been performed to study the pick-up of a singly charged Fe(CO)$_5^+$ molecule on the cluster at the same collision parameters. Such information can be useful e.g. for experiments using ions in argon matrices \cite{jacox02}.
As shown in Figure~\ref{fig:pickup}(b), there is a minor difference in penetration of the neutral and ionic iron pentacarbonyl molecules into the argon cluster, and the molecule's charge state has a minor impact on the average distance between the iron atom and the center of mass of the cluster. A more detailed and systematic analysis of the geometry of Fe(CO)$_5^+$ inside argon clusters as a function of collision parameters goes beyond the scope of the present study.

\begin{figure}[tb]
\includegraphics[width=0.44\textwidth]{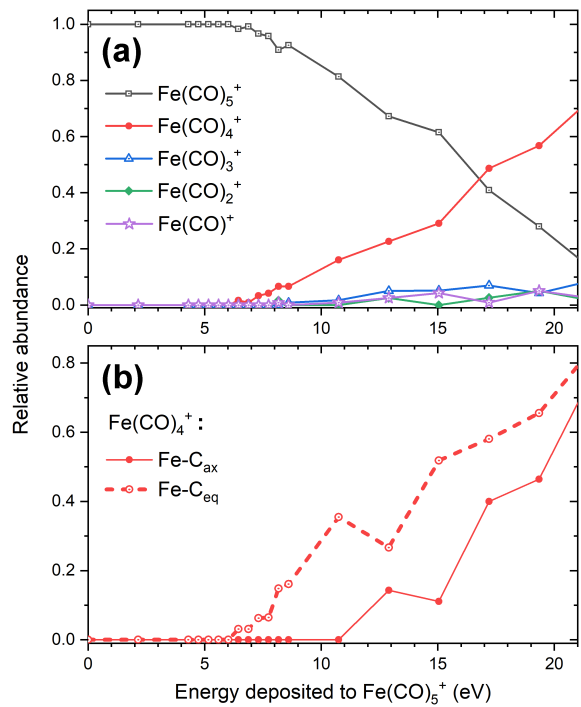}
\caption{Panel \textbf{(a)}: Relative abundance of Fe(CO)$_{5-n}^+$ ($n=0-4$) ionic species produced due to the electron impact induced fragmentation of a Fe(CO)$_5^+$ molecule embedded into the argon cluster. The excess energy was deposited locally into one of the C--O bonds of the molecule. Panel~\textbf{(b)}: Relative abundance of Fe(CO)$_4^+$ fragments. Solid line corresponds to the localized energy deposition in an axial (Fe--C$_{\rm ax}$ coordinated) C--O bond; dashed line describes the case of energy deposition in an equatorial (Fe--C$_{\rm eq}$ coordinated) C--O bond.}
\label{fig:dist_ar}
\end{figure}

Figure~\ref{fig:dist_ar}(a) shows the calculated relative abundances of Fe(CO)$_{5-n}^+$ ($n=0-4$) ionic species produced due to the dissociation of a Fe(CO)$_5^+$ molecule embedded into the argon cluster. The results differ significantly from those obtained in the gas phase (see Fig.~\ref{fig:gas_phase}). In the case of Fe(CO)$_5^+$@Ar, most of the Fe(CO)$_5^+$ ions have remained intact at the end of 25~ns long simulations (black curve with open squares), and the parent Fe(CO)$_5^+$ ion dominates the spectrum up to the deposited energies of $E \sim 17$~eV. The only ionic fragment with non-negligible abundance is Fe(CO)$_4^+$ (red curve with circles), corresponding to the loss of one CO ligand. We note that in the simulations of Fe(CO)$_5^+$@Ar fragmentation, all the excess energy has been deposited locally into one of the C--O bonds. The cause of the ligand stabilization is a fast energy transfer from the dissociating bond to the argon environment, which serves as a heat bath that efficiently quenches the excess energy from the Fe(CO)$_5^+$ cation. Most dissociation events observed in the simulations have occurred on a timescale shorter than 100~ps, indicating a prompt fragmentation. The energy deposited into a C--O bond has been transferred into the vibrations of Fe--C bonds, eventually leading to the breakage of the metal--ligand bonds. This process occurs before the energy is quenched by the argon environment.

The observed fragmentation pattern has an interesting consequence. Figure~\ref{fig:dist_ar}(b) shows the Fe(CO)$_4^+$ abundance for the case of the localized energy deposition into the axial or equatorial C--O bonds. The appearance energies for the Fe(CO)$_4^+$ fragment differ in these two cases by more than 4~eV. Therefore, one concludes that the difference in the strength of the metal--ligand bonds ($\sim$0.5~eV) has a strong impact on the efficiency of the energy transfer to the argon environment. This result is in strong contrast to the case of an isolated molecule, where the dissociation was not prompt but thermally driven, and the difference between the two types of bonds was much smaller (see Figure~\ref{fig:gas_phase}(d)).

An important observation from Ref.~\citenum{Lengyel2016_JPCC_2} should also be mentioned here in connection to the present simulation outcomes. The measurements of appearance energies revealed that a substantial fraction of Fe(CO)$_5$ molecules on the cluster are not ionized by direct electron impact, but rather the electron ionizes argon atoms in the cluster, and Fe(CO)$_5$ is then ionized by a hole transfer from argon \cite{Lengyel2016_JPCC_2}. This was concluded based on the fact that the appearance energies of fragments coincided with the ionization energy of an argon atom. This experimental observation is supported by the results shown in Fig.~\ref{fig:pickup} that the Fe(CO)$_5$ molecule is embedded into the cluster after the pick-up. Considering the ionization potentials of an argon atom (15.8~eV) \cite{183436} and a Fe(CO)$_5$ molecule (8.45~eV) \cite{Lacko_EPJD2015}, such charge transfer is exothermic by $\sim$7.3~eV. As demonstrated in Figure~\ref{fig:dist_ar}(a), almost all ions formed in this way remain intact and do not experience fragmentation.

\begin{figure}[t]
\includegraphics[width=0.46\textwidth]{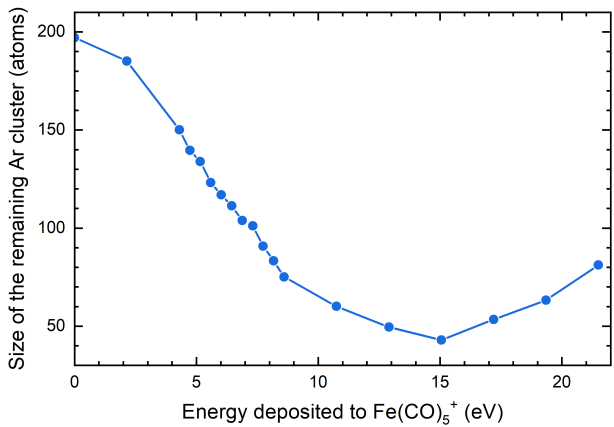}
\caption{The average number of argon atoms in the cluster surrounding the Fe(CO)$_5^+$ ion at the end of the 25-ns long simulation as a function of energy deposited into the Fe(CO)$_5^+$ molecule.}
\label{fig:cluster_size}
\end{figure}

The energy transferred to the cluster is redistributed among the internal degrees of freedom of the cluster, which leads to the evaporation of the weakly bound argon atoms. Figure~\ref{fig:cluster_size} shows the remaining number of argon atoms in the cluster at the end of the 25~ns long simulations as a function of energy deposited into the Fe(CO)$_5^+$ molecule, $E$. As the deposited energy increases, the resulting cluster size decreases, which is expected. However, this trend breaks at the deposited energy of $E \sim 15$~eV. The main reason for this phenomenon is that for larger values $E$, a larger number of CO ligands are released by a prompt dissociation with a small energy loss to argon atoms, yielding mainly Fe(CO)$_4^+$ as demonstrated in Figure~\ref{fig:dist_ar}(a).

\section{Conclusions}
\label{sec:Conclusions}

In conclusion, the dissociative ionization (DI) of the iron pentacarbonyl molecule, Fe(CO)$_5$, has been studied by means of reactive molecular dynamics simulations using the MBN Explorer software package \cite{Solovyov_2012_JCC_MBNExplorer}. The main focus of this study concerned the quantitative analysis of different ionic fragments and their appearance energies for the fragmentation of a single Fe(CO)$_5^+$ molecule in the gas phase and the molecule embedded into a molecular environment.

For an isolated Fe(CO)$_5^+$ molecule in the gas phase, different fragmentation mechanisms corresponding to the distribution of excess energy into different covalent bonds have been studied. The main outcome of the simulations -- abundances of individual fragments -- show a surprisingly little dependence on the initial conditions. This observation is attributed to intramolecular vibrational redistribution (IVR), which means that on a short timescale, the excess energy becomes distributed over the internal degrees of freedom of the whole molecule, and the dissociation of metal--ligand bonds in Fe(CO)$_5^+$ proceeds via the thermal mechanism of fragmentation. The evaluated appearance energies of individual fragments are in very good agreement with experimental values \cite{Lacko_EPJD2015}.

In the case of iron pentacarbonyl embedded into an argon cluster, the release of CO ligands is strongly suppressed. The simulation results reported in this study provide an atomistic understanding of the cluster-beam study of Lengyel et al.~\cite{Lengyel2016_JPCC_2}, who observed such stabilization experimentally. We have demonstrated that the excess energy deposited to the Fe(CO)$_5^+$ cation as a result of the electron collision is efficiently quenched by the argon environment, which leads to the heating of the cluster and the evaporation of weakly bound argon atoms. The simulations carried out in this study also bring experimentally inaccessible information about the structure of the heterogeneous Fe(CO)$_5^+$@Ar cluster following the pick-up collision of  Fe(CO)$_5$ molecules with pristine argon clusters. It has been demonstrated that Fe(CO)$_5$ is embedded into the argon cluster, and the penetration depth of the picked-up molecule does not depend on its charge state (a neutral or a singly charged positive species).

The present findings are relevant for understanding the irradiation-driven fragmentation of molecular systems placed in molecular environments. Several effects related to a molecular environment and influencing the fragmentation degree have been discussed in the literature, such as mechanical caging \cite{fedor_hbr11}, stabilization of transient species \cite{chachereau16}, change of chemical pathways \cite{kocisek_dCMP18}, or polarization effects \cite{fabrikant_review18, Lengyel2017_Beilstein}. The present findings elucidate in a quantitative way one of the most common effects of the molecular environment  -- quenching of the excess energy deposited into the system during the fragmentation process.

The simulations carried out in this study are also relevant to the question of the DI of precursor molecules during the focused electron beam induced deposition (FEBID) process. Even though the relative contribution of DI (with respect to the dissociative electron attachment and neutral dissociation processes) varies with the energy of secondary electrons in a given deposition process, this contribution is always significant~\cite{Thorman2015, zlatar16}. The efficient IVR mechanism observed and analyzed in this study indicates that the iron pentacarbonyl cations are almost always vibrationally excited, and the loss of CO ligands is a thermally driven process. When the molecules are physisorbed on a substrate -- as in the case of FEBID -- the vibrational energy can be efficiently quenched by the environment. The process of quenching is similar to that observed here by the argon cluster. 
Understanding irradiation-driven fragmentation patterns for molecular systems in environments facilitates the advancement of advanced computational models for studying irradiation-induced chemistry processes involving complex molecular systems \cite{Sushko2016_IDMD, DeVera_2020_SciRep, Prosvetov2021_BJN, Prosvetov2022_PCCP}.

\begin{acknowledgments}

This work has been supported by the European Union's Horizon 2020 research and innovation programme – the RADON project (GA 872494) within the H2020-MSCA-RISE-2019 call; the COST Action CA20129 MultIChem supported by COST (European Cooperation in Science and Technology); the Czech Science Foundation project 20-11460S, and by European Regional Development Fund, OP RDE, project “CARAT” no. CZ.02.1.01/0.0/0.0/16\_026/0008382. The possibility of performing computer simulations at the Goethe-HLR cluster of the Frankfurt Center for Scientific Computing is also gratefully acknowledged.
\end{acknowledgments}

\bibliography{FeCO5_references}

\end{document}